**Metallic and Insulating Adsorbates on Graphene**

K. M. McCreary, K. Pi,[†] and R. K. Kawakami[‡]

Department of Physics and Astronomy, University of California, Riverside, CA 92521

**Abstract**

We directly compare the effect of metallic titanium (Ti) and insulating titanium dioxide ($TiO_2$) on the transport properties of single layer graphene. The deposition of Ti results in substantial *n*-type doping and a reduction of graphene mobility by charged impurity scattering. Subsequent exposure to oxygen largely reduces the doping and scattering by converting Ti into $TiO_2$. In addition, we observe evidence for short-range scattering by $TiO_2$ impurities. These results illustrate the contrasting scattering mechanisms for identical spatial distributions of metallic and insulating adsorbates.

[†]Present address: Hitachi Global Storage Technologies, San Jose, CA 95135

[‡]e-mail: roland.kawakami@ucr.edu

The interaction of electrons in graphene with surface adsorbates is a central issue for electronic mobility, correlated electron physics[1, 2], and applications in chemical sensors[3, 4]. Furthermore, metals and insulators on graphene surfaces are employed for essential device elements such as electrical contacts[5], gate dielectrics[6, 7], and tunnel barriers[8, 9], so it is important to understand their influence on the electronic properties of graphene. While experiments have separately investigated the doping of graphene by metallic[10-12] and insulating adsorbates[12, 13], the transition from metallic to insulating behavior of the adsorbates has not been explored.

In this paper, we compare the electronic properties of graphene doped with metallic Ti and insulating $TiO_2$ and investigate the transition between the metallic and insulating states of the adsorbates. The experiments are performed in ultrahigh vacuum (UHV) where submonolayer amounts of Ti are deposited onto graphene electronic devices and subsequently converted to $TiO_2$ by a partial pressure of oxygen gas. The evolution of the electronic properties is monitored by *in situ* transport measurements during the entire deposition and oxidation process, which allows direct comparison between Ti- and $TiO_2$-doping for identical spatial distribution of the dopant atoms. We find that the Ti doping produces substantial charge transfer to the graphene (*n*-type doping) and a strong reduction of mobility due mainly to charged impurity scattering. As the Ti is converted to $TiO_2$, the effects of the adsorbates are largely reversed: both the charge transfer and impurity scattering are greatly reduced. This clearly demonstrates that insulating adsorbates have much less influence on the electronic properties of graphene as compared to metallic adsorbates. Furthermore, we find evidence that $TiO_2$ impurities produce short-range scattering in single layer graphene. Because Ti is often used for adhesion of electrical contacts and $TiO_2$ is used as a seed layer for tunnel barriers[8], the results for these materials are particularly important for understanding the electronic and spintronic properties of graphene devices.

Samples are prepared by mechanical exfoliation of highly oriented pyrolytic graphite onto a $SiO_2$/Si substrate (300 nm thickness of $SiO_2$). Single layer graphene (SLG) flakes are identified by optical microscopy and Raman spectroscopy[14]. Ti/Au electrodes are defined by electron beam lithography. Devices are then annealed in a Ar/$H_2$ environment to remove resist residue [15, 16] and loaded into UHV to be degassed[11].

The sample is then transferred under UHV to a molecular beam epitaxy (MBE) system where *in situ* transport measurements are performed during the deposition of Ti adatoms. The Ti is deposited from a triode electron beam source[17]. During the Ti deposition, the pressure remains below $7\times10^{-10}$ torr. The sample is held at room temperature, which will promote the formation of clusters[12]. The deposition rate is calibrated by a quartz deposition monitor. The submonolayer coverage is converted from atoms/$cm^2$ to "monolayers" (ML) where 1 ML is defined as $1.908\times10^{15}$ atoms/$cm^2$, the areal density of primitive unit cells in graphene.

Figure 1(a) displays the gate dependent conductivity ($\sigma$) for clean graphene and for select amounts of Ti deposition. The conductivity is measured in a four-probe geometry using standard lock-in detection techniques. The minimum conductivity ($\sigma_{min}$) as a function of gate voltage ($V_g$) identifies the Dirac point ($V_D$). The carrier concentration is calculated as $n = -(\kappa\varepsilon_0/ed)(V_g - V_D)$, where $\kappa$ is the relative dielectric constant of $SiO_2$ (3.9), $\varepsilon_0$ is the permittivity of free space, and $d$ is the dielectric thickness (300 nm). This yields $n = -\alpha(V_g - V_D)$ with $\alpha = 7.2\times10^{10}$ $V^{-1}cm^{-2}$. The slope of $\sigma$ in the linear regions to the right (left) of the $V_D$ is directly related to electron (hole) mobility through the relation $\mu_{e,h}=\Delta\sigma/e\Delta n$. Prior to Ti deposition, the clean graphene exhibits a $V_D$ close to zero $V_g$, indicating little chemical doping. The average mobility, defined as $\mu=(\mu_e+\mu_h)/2$, is 4830 cm$^2$/Vs, typical for graphene supported by a SiO$_2$ substrate. As Ti is deposited, V$_D$ shifts to more negative gate voltages, indicating electrons are being transferred from Ti to graphene. The charge transfer results in positively charged impurities on the surface, which act as sources of additional scattering for charge carriers in the graphene. The widening of the curves in Figure 1(a) with increased Ti deposition signifies the mobility is decreasing, and is a clear indication of the additional scattering. Figure 1(b) plots the values for the V$_D$, while Figure 1(c) shows the average mobility for the clean sample and five Ti depositions. As a result of 0.028 ML Ti deposition, V$_D$ shifts from +3.5 V to -42.5 V while μ is reduced from 4830 to 1390 cm$^2$/Vs, consistent with previous studies[11].

Following the deposition of Ti, the sample is left undisturbed in the UHV environment while the conductivity is measured at discrete time intervals. Figure 2(a) shows the gate dependent conductivity measured immediately after deposition along with 6, 23, 47, and 103 minutes after the 0.028 ML Ti deposition. Note that the data labeled 0 min in Figure 2(a) is labeled 0.028 ML Ti in Figure 1(a). During this time, there are small changes in both mobility and V$_D$, most likely due to cluster formation[12]. Subsequently, oxygen is leaked into the chamber at a pressure of 1×10$^{-6}$ torr. Exposure to oxygen has an immediate and drastic effect on the measured transport properties. The Dirac point shifts to more positive voltages, showing that the transformation from metallic Ti to insulating TiO$_2$ decreases the electronic doping level of the graphene. This is reasonable when one compares metals and insulators on a simplified level. Metals have free electrons which can easily be transferred to graphene, whereas the electrons in insulators are tightly bound. Therefore the presence of TiO$_2$ on graphene should have little effect on the electronic doping. In conjunction with the decrease in doping is an increase in mobility, as is seen in Figure 2(b) by the narrowing of the curves with longer oxygen exposure. The results of Ti deposition and oxidation are summarized in Figure 3, which shows the gate dependent conductivity for the clean device, 0.028 ML Ti and fully oxidized Ti impurities. The final Dirac point of 0 V almost fully recovers to the initial Dirac point of +3.5 V. To ensure that these effects are not related to physisorption of oxygen to the graphene surface (which could produce electronic doping[18]), we perform control experiments on clean graphene devices in UHV. For an exposure of 1300 Langmuirs of $O_2$, there is no observable change in the mobility and the Dirac point shifts by less than 2 V, indicating that any physisorbed oxygen has negligible effect.

The recovery of the Dirac point to nearly its initial value relies on the property of autocompensation of $TiO_2$ surfaces[19]. The strong electronegativity of oxygen results in Ti being positively charged while the oxygen will be negatively charged. In principle, this could produce a polar surface that introduces electronic doping in the graphene: if the $TiO_2$ is terminated by Ti, the positive bound charges would induce a strong *n*-type doping in the graphene. However, UHV studies find that $TiO_2$ surfaces are autocompensated[19], meaning the proportion of Ti and O surface atoms are balanced, making the surface charge-neutral. The incomplete recovery of the Dirac point suggests that there is not a perfect autocompensation between the Ti and O—the chemical interaction between the Ti atoms and the graphene may alter the energetics of autocompensation.

To investigate the relationship between electron scattering and the amount of charge on the impurities, we plot the mobility and minimum conductivity as a function of the Dirac point voltage. Figure 4(a) shows μ vs. $V_D$ and figure 4(b) shows $\sigma_{min}$ vs. $V_D$ throughout the Ti doping, cluster formation, and oxidation steps. The effect of Ti doping is represented by the solid black circles and follows the path from "1" to "2" in Figures 4(a) and 4(b) with $V_D$ shifting to negative values while both μ and $\sigma_{min}$ are reduced. The cluster formation is represented by green crosses from "2" to "3". The oxidation is represented by the open red circles from "3" to "4" with $V_D$ shifting back toward zero while μ and $σ_{min}$ are both increasing. Interestingly, the paths followed by μ and $σ_{min}$ during the oxidation step ("3" to "4") fall on nearly the same curves as for the initial Ti deposition ("1" to "2"). The overlap of the deposition and oxidation curves is expected if the main scattering mechanism is long-range charged impurity scattering. In this case the electron scattering should depend only on the amount of charge located on the impurities (i.e. the Dirac point voltage) and not on its chemical species or local bonding. Thus, these results indicate that the main scattering mechanism for Ti and partially oxidized Ti is charged impurity scattering. Looking more closely, the oxidation curves for $σ_{min}$ and μ exhibit a different behavior when the Ti is nearly fully oxidized (near "4"). When the oxidation curves are extrapolated to the initial $V_D$ of +3.5 V, $σ_{min}$ extrapolates to the initial value for clean graphene ("1", Fig 4(b)), but μ extrapolates to a value less than the initial value for clean graphene ("1", Fig 4(a)). The different behavior occurs because $σ_{min}$ is determined by the formation of electron-hole puddles which are generated by the long-range charged impurity scattering[13, 20]. On the other hand, μ can have contributions from both long-range and short-range scattering. Thus, $σ_{min}$ is expected to have a full recovery for charge-neutral impurities, while the reduced value of μ (extrapolated to $V_D$ → +3.5 V) indicates a contribution from short-range scattering by $TiO_2$ impurities. This is consistent with recent evidence for short-range scattering by insulating layers on SLG[7].

In conclusion, we have monitored the electronic properties of doped graphene as impurities are transformed from Ti to $TiO_2$. It is determined that the conversion of adsorbates from metallic to insulating behavior largely reverses both the charge transfer and charged impurity scattering. The trends observed in mobility indicate that the presence of $TiO_2$ introduces additional short-range scattering to the system.

**Acknowledgements**

We acknowledge technical assistance and discussion with W. Han, A. Swartz, J. Wong, and H. Wen, and the support of ONR (N00014-09-1-0117), NSF (DMR-1007057), and NSF (MRSEC DMR-0820414).

**Figure Captions**

Figure 1: (a) The gate dependent conductivity at selected values of Ti coverage. (b, c) The Dirac point and mobility, respectively, as a function of Ti coverage.

Figure 2: (a) The gate dependent conductivity is displayed for 0, 6, 23, 47 and 103 minutes following the final Ti deposition. (b) The gate dependent conductivity prior to $O_2$ exposure and after 1, 14, 34, 78, 203, and 1503 minutes of exposure to oxygen at a pressure of $1\times10^{-6}$ torr.

Figure 3: A comparison of the gate dependent conductivity curves measured for the clean sample, after 0.028 ML Ti deposition, and after 1503 minutes of oxygen exposure.

Figure 4: (a, b) The mobility and minimum conductivity, respectively, as a function of gate voltage. The solid black circles in each graph correspond to Ti deposition, green crosses to cluster formation, and open red circles to oxidation.

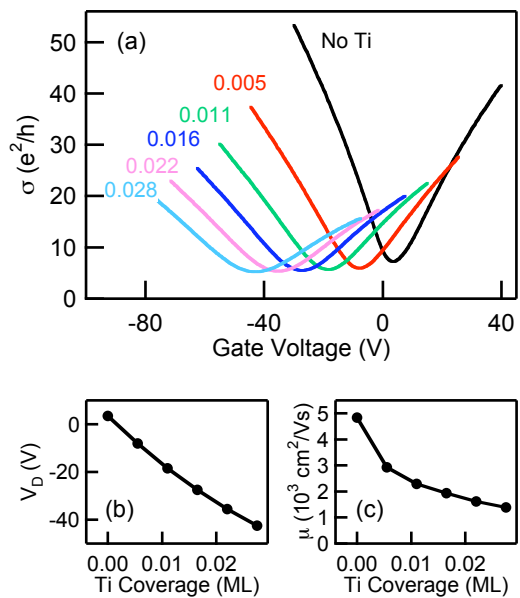

Figure 1

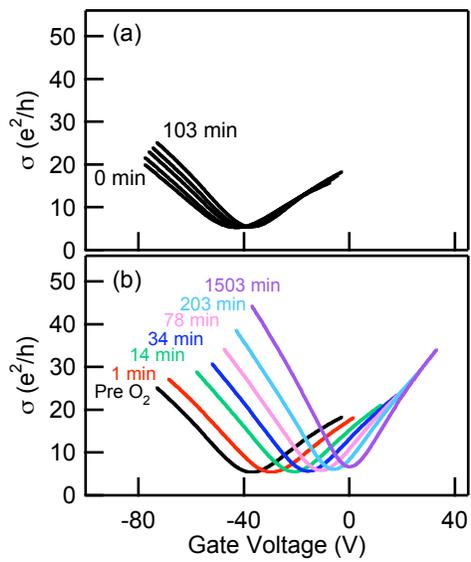

Figure 2

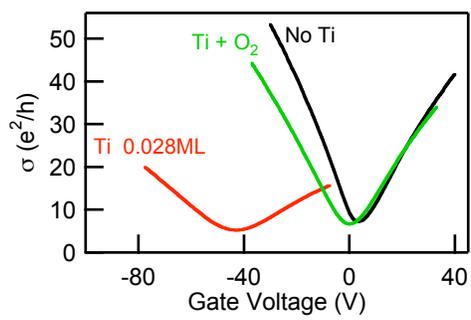

Figure 3

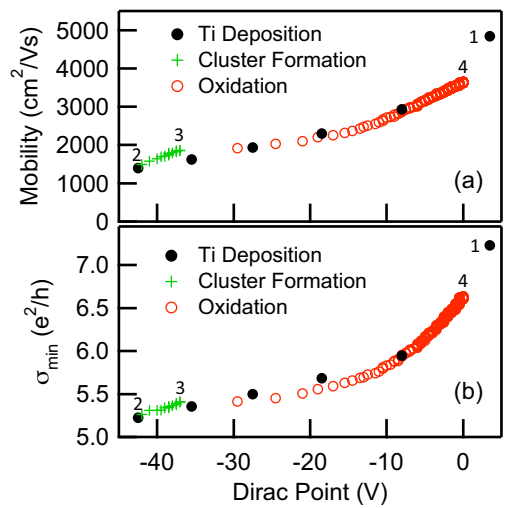

Figure 4